\documentclass[12pt,letterpaper]{JHEP3}
\usepackage{amsmath}
\usepackage{graphicx}
\usepackage{amsfonts}
\usepackage{amssymb}
\usepackage{epsfig}
\usepackage{psfrag}

\newcommand{\be}{\begin{equation}}
\newcommand{\ee}{\end{equation}}
\newcommand{\beq}{\begin{eqnarray}}
\newcommand{\eeq}{\end{eqnarray}}

\title{Do many-particle neutrino interactions cause a novel coherent effect?}

\author{Alexander Friedland\\
Theoretical Division, T-8, MS B285, Los Alamos National Laboratory, Los Alamos, NM 87545 \\
E-mail: \email{friedland@lanl.gov}}
\author{Cecilia Lunardini\\
School of Natural Sciences, Institute for Advanced Study, Einstein Drive, Princeton, NJ 08540 \\
E-mail: \email{lunardi@ias.edu}}

\abstract{We investigate whether coherent flavor conversion of
neutrinos in a neutrino background is substantially modified by
many-body effects, with respect to the conventional one-particle
effective description. We study the evolution of a system of
interacting neutrino plane waves in a box. Using its equivalence
to a system of spins, we determine the character of its behavior
completely analytically. We find that, if the neutrinos are
initially in flavor eigenstates, no coherent flavor conversion is
realized, in agreement with the effective one-particle
description.  This result does not depend on the size of the
neutrino wavepackets and therefore has a general character. The
validity of the several important applications of the one-particle
formalism is thus confirmed. }

\preprint{hep-ph/0307140, LA-UR-03-3847}

\begin{document}

\section{Introduction}

The flavor composition of a neutrino system may be modified
as a result of the interactions between the neutrinos and a
background medium. The efficiency of this process depends, in
addition to the properties of the background itself (composition,
density, temperature, etc.), on the  coherent (or
incoherent) character of the interaction.

Here we study the case of \emph{coherent} neutrino conversion in a
background of {\it neutrinos}. The interaction of a neutrino beam
with a neutrino background has the peculiar feature that, due to
momentum exchange, a neutrino from the background may be scattered
into the beam and vice versa.  This differs from the case of
scattering on electrons and/or nucleons, and implies that the
effect of coherent scattering on neutrinos can not be treated in
analogy with that on ordinary matter.

In the literature, the effects of the neutrino-neutrino coherent
scattering on neutrino conversion have traditionally been treated
following the pioneering work by Pantaleone
\cite{Pantaleone1992prd,Pantaleone1992plb}, Sigl \& Raffelt
\cite{germanpaper}, and McKellar \& Thomson \cite{McKellarThomson}. In
these papers, a \emph{single-particle} evolution equation for each
neutrino is derived, which has the form of the usual oscillation
equation for neutrinos in matter with an extra term designed to
capture the effect of neutrino--neutrino scattering.  According to this
equation, the effect of the interaction with the
neutrino background is significant if the neutrino-neutrino potential
is at least comparable to the potential due to matter
(zero-temperature and thermal terms) and/or to the vacuum oscillation
terms:
\begin{equation}\label{eq:criterion}
    |n_\nu-n_{\bar\nu}|\gtrsim\left\{
    \begin{array}{c}
    |n_{e^-}-n_{e^+}|\\
    (n_{e^-}+n_{e^+})E_\nu^2/m_w^2\\
    \Delta m^2/(G_F E_\nu)
    \end{array}\right.
\end{equation}
Here $n_\nu$  $(n_{\bar\nu})$ is the (anti)neutrino number
density, $n_{e^-}$ $(n_{e^+})$ is the electron (positron) number
density, $E_\nu$ is the neutrino energy, $m_w$ is the $W$-boson
mass and $G_F$ is the Fermi constant.

The condition (\ref{eq:criterion}) may be satisfied in the early
Universe close to the Big Bang Nucleosynthesis epoch (at
temperature $T\sim O(1)$ MeV) or just outside a supernova core (at
$r\sim O(100{\rm\; km})$).  Indeed, in these environments the
number density of neutrinos is comparable to, or larger than, that of
electrons and nucleons.  At the same time, incoherent scattering is
negligible and neutrinos stream freely in the medium, affected by
coherent scattering (refraction) only. The evolution of the
neutrino flavor composition in both cases has been extensively
studied using the one-particle equation. An important result
\cite{Lunardini2000,Hannestad:2001iy,Dolgovetal2002, Wong2002,
Abazajian2002} is the strong bound on the chemical potential of
non-electron relic neutrinos that follows from studying neutrino
oscillations in the early Universe, with important implications on
BBN.
Other important applications include hypothetical oscillations
between active and sterile neutrino states that may have generated
a lepton asymmetry in the early Universe
\cite{FootVolkas1996,FootVolkas1997,BellFootVolkas1998,Dolgov1999,DiBarietal2000}
and possible effects of a sterile neutrino on the $r$-processes in a
supernova
\cite{Pantaleone1995,QianFuller1995,Sigl1995,McLaughlin1999,PastorRaffelt2002}.
\\

There is an important theoretical issue at the foundation of all of
the mentioned analyses that, if unresolved, casts doubt on their
validity. The issue is the \emph{existence and the range of validity}
of the description of neutrino self-refraction in terms of a set of
one-particle equations. Generally speaking, the interactions between
neutrinos in the ensemble may be expected to create quantum
correlations, or, in other words ``entangled'' states of many
neutrinos which are not products of individual wavefunctions.  \emph{A
  priori}, a system in which such entanglement exists requires a
many-particle description. The importance of this point was recognized
early on \cite{Pantaleone1992plb}, but was not pursued further in the
following years.

Recently, the problem of possible existence and effects of quantum
correlations has been investigated using two different approaches.
The first, put forth in our recent paper \cite{us}, is to understand
the flavor evolution of a many-neutrinos system as an effect of the
interference of many elementary neutrino-neutrino scattering events.
Using this construction, we argued that, in the limit of infinite
number of neutrinos, the many-body description factorizes into
one-particle equations.  These equations coincide with those given in
the previous literature up to terms which change the overall phases of
neutrino states. While such terms can be potentially important in
non-oscillation phenomena, they do not affect the flavor evolution of
the system.

As a particular example, we considered an ensemble of neutrinos which
are initially in flavor eigenstates. Following our analysis, we found
that in this system, somewhat counterintuitively, no coherent flavor
conversion takes place. This finding is in agreement with the standard
one-particle description of
\cite{Pantaleone1992prd,Pantaleone1992plb,germanpaper,McKellarThomson}.

The second approach, introduced by Bell {\it et al.}
\cite{Belletal}, is to consider the evolution of a system of
interacting neutrino plane waves in a box. The idea is to study
the properties of the many-body Hamiltonian of this system with
the goal of determining the rate with which statistical
equilibration is achieved.
The analysis is applied 
to a system of neutrinos which are initially in flavor
eigenstates. It is concluded that the time scale $t_{eq}$ of the
flavor evolution of a given neutrino in this system is inversely
proportional to the neutrino density and to the Fermi constant:
$t_{eq}\propto (n G_F)^{-1}$, which is characteristic of coherent
flavor conversion.
This result contrasts with the prediction of the one-particle
description, and therefore has been considered as an indication of
its breakdown and of the existence of new coherent effects due to
entanglement. Clearly, this would have profound implications on
the applications of neutrino-induced neutrino conversion to
supernova physics and cosmology.
\\

Given the importance of the subject, in this paper we present a
further study of neutrino flavor conversion in a neutrino
background. The aim is to clarify the origin of the contrast
between Ref. \cite{Belletal} and the other literature and give a
definite answer to the question of the importance of entanglement
and its effects.

Since the suggested new effect comes from simultaneous interaction
between many neutrinos, we study, following \cite{Belletal}, the
Schr\"odinger equation of a system of many neutrino plane waves in
a box.  We consider the free streaming regime and neglect vacuum
oscillation terms, which can be included by a straightforward
generalization of our results. We show that the problem is
equivalent to that of a system of spins, for which, in the case of
constant spin-spin coupling, the equilibration time can be
determined completely analytically. For neutrinos initially in
flavor eigenstates, in the limit of the infinite number of
neutrinos, this solution shows no coherent conversion, in
agreement with our earlier results.

The paper is organized as follows. In Sect. \ref{sect:prelim} we lay
the foundations of our analysis, by discussing the structure of the
neutrino--neutrino interaction in the flavor space and in the real
space. We also perform an elementary analysis of coherence in a system
of several interacting neutrinos and use it to motivate the following
study. In Sect. \ref{sect:twoapproaches} the many body approaches are
discussed and the relevant results of our previous paper are reviewed.
In Sect. \ref{sec:full} we formulate the problem of interacting
neutrino plane waves in a box, present the full analytical solution
and give its derivation. A discussion and conclusions follow in Sect.
\ref{sec:disc} and \ref{sec:conclusions}.

\section{General considerations}
\label{sect:prelim}


\subsection{The flavor structure of the interaction}
\label{subsect:flavor}

In the range of neutrino energies that are relevant to the
applications ($\sim 10^6-10^7$ eV), the interaction between neutrinos
is described by the low-energy neutral current (NC) Hamiltonian
\begin{eqnarray}
  \label{eq:NChamiltonian}
  H_{\rm NC}= \frac{G_F}{\sqrt{2}}
\left(\sum_a \bar\nu_a\gamma^\mu\nu_a\right)
\left(\sum_b\bar\nu_b\gamma_\mu\nu_b\right).
\end{eqnarray}
Higher order effects, such as those coming from the expansion of
the propagator or particle production (bremsstrahlung) will be
neglected in our analysis.

A property of this Hamiltonian that will play a very important
role later is its invariance under the rotation of the $SU(2)$
flavor group. Because of this property, the interaction between
pairs of neutrinos -- viewed in flavor space -- must be equivalent
to the interaction between pairs of spins.

The one-particle equations of
\cite{Pantaleone1992prd,Pantaleone1992plb,germanpaper,McKellarThomson,us}
indeed preserve the $SU(2)$ structure of the problem and have a
form of the interaction between spins\footnote{In
\cite{Pastor:2001}, this property of the equations has been used
to give an elegant physical analogy designed to explain the
``synchronized" oscillations of neutrinos in a neutrino
background.}.
It would be logically inconsistent, however, to rely on this fact
in the analysis, in which these equations are being tested.
Therefore, below we present an explicit proof of the equivalence,
valid regardless of whether the interaction is coherent or not.

Let us consider the interaction of two neutrinos, and, in the
interests of clarity, for a moment suppress the Dirac indices on
the fermions and the gamma matrices. Since the NC interactions
conserve flavor, in the Hamiltonian (\ref{eq:NChamiltonian}) the
flavor space wavefunction of a given outgoing neutrino is equal to
that of one of the incoming neutrinos. Let us denote the flavor
wavefunctions by $\Psi_i$ and $\Phi_i$. There are then two
possible combinations:
\begin{eqnarray}
  \label{eq:flavoroption1}
  \Psi_i^\ast\delta_{ij}\Psi_j \Phi_k^\ast\delta_{kl}\Phi_l,\\
  \label{eq:flavoroption2}
  \Psi_i^\ast\delta_{il}\Phi_l \Phi_k^\ast\delta_{kj}\Psi_j.
\end{eqnarray}

Since $\sum_{i=1}^{2}|\Psi_i|^2=\sum_{i=1}^{2}|\Phi_i|^2=1$, the term
in Eq.~(\ref{eq:flavoroption1}) equals $1$. The term in
Eq.~(\ref{eq:flavoroption2}) could be transformed using the well-known
property of the (4-dimensional) $\sigma$-matrices,
$\sigma_{\alpha\dot\alpha}^m\bar\sigma_m^{\dot\beta\beta}=-2\delta_{\alpha}^\beta\delta_{\dot\alpha}^{\dot\beta}$
(in the notation convention of \cite{WessBagger}), or
\begin{eqnarray}
  \label{eq:sigmatrick}
  2 \delta_{il}\delta_{jk}=
  \delta_{ij}\delta_{kl}+\vec\sigma_{ij}\cdot\vec\sigma_{kl}.
\end{eqnarray}
One gets
\begin{eqnarray}
  \label{eq:flavortransform}
  \Psi_i^\ast\delta_{il}\Phi_l \Phi_j^\ast\delta_{jk}\Psi_k=
  \frac{1}{2}(1+
  \Psi_i^\ast\vec\sigma_{ij}\Psi_j \cdot\Phi_k^\ast\vec\sigma_{kl}\Phi_l).
\end{eqnarray}
In this form, the equivalence between a system of neutrinos and a
system of spins is manifest. The complete (including both the
contributions from (\ref{eq:flavortransform}) and
(\ref{eq:flavoroption1})) flavor space Hamiltonian for the
interaction of two neutrinos, 1 and 2, is proportional to
\begin{eqnarray}
  \label{eq:spinH}
    (3/2+\vec\sigma_1\cdot\vec\sigma_2/2),
\end{eqnarray}
which coincides with the square of the operator of the \emph{total}
angular momentum, $\hat{L}^2\equiv (\vec\sigma_1/2+\vec\sigma_2/2)^2$,
as expected.

The strength of the interaction depends on the scattering angle in
real space. Let us denote the states of the incoming neutrinos as
$\chi_{\alpha}$ and $\psi_{\beta}$ and those of the outgoing
states as $\bar\psi_{\dot\beta}'$ and $\bar\chi_{\dot\alpha}'$.
The spatial dependence of the scattering amplitude is given by
\begin{equation}\label{eq:gen_angle}
    \bar\psi_{\dot\beta}'\bar\sigma^{\dot\beta\beta m} \psi_{\beta}
    \bar\chi_{\dot\alpha}'\bar\sigma^{\dot\alpha\alpha}_{m}
    \chi_{\alpha}= 1-
    (\bar\psi'\vec\sigma \psi)\cdot
    (\bar\chi'\vec\sigma \chi)
\end{equation}

If the neutrino wavepackets are sufficiently broad so that several
wavepackets overlap, the preceding argument can be trivially
generalized to show that the interactions between the
``overlapping" neutrinos have the structure of the interactions
between the corresponding spins. In the case of neutrino plane
waves in a box, each spin interacts with all the others and the
total Hamiltonian is the sum of the interactions of all pairs (see
Sect. \ref{sec:der}).

\subsection{Coherence of neutrino-neutrino scattering: basic features}
\label{subsect:cohvsincoh}

We next discuss general properties of coherent
neutrino-neutrino scattering.

By definition, scattering is coherent when the waves scattered by
different particles in a target interfere with each other. When
the scatterer is distinguishable from the incident particle, the
coherence condition is satisfied when the incident particle is
scattered forward. For other directions, scattering is incoherent
(unless the particle spacings in the target satisfy particular
conditions, e.g. periodicity).
In contrast, for neutrino-neutrino scattering, scattering
can be coherent when the momentum of the scattered neutrino coincides
with the momentum of \emph{one} of the incident neutrinos.  The
corresponding values of the scattering angle $\beta$ are $0$ and
$\Theta$, where $\Theta$ denotes the angle between the two
initial-state neutrinos.

For the purpose of studying coherent effects, one may replace the
full Hamiltonian (\ref{eq:NChamiltonian}) with a ``toy"
Hamiltonian which restricts scattering angles to values $0$ and
$\Theta$ only, for which coherent interference may occur. This
simplified interaction can be viewed as a flavor exchange between
neutrinos which do not change their momenta. We emphasize that,
while all coherent effects in the system are fully
captured in this way, most of the incoherent effects are
left out. Hence, one must be careful while interpreting incoherent
effects found in this framework, as will be seen later.

Following \cite{Pantaleone1992plb}, we write the reduced
Hamiltonian for a pair of interacting neutrinos with momenta
$\vec{k}$ and $\vec{p}$ in the form
\begin{eqnarray}
  \label{eq:pant4by4}
  i\frac{d}{dt}\left(
    \begin{array}{c}
      |\nu_e (\vec{k}) \nu_e(\vec{p})\rangle  \\
      |\nu_e (\vec{k}) \nu_\mu(\vec{p})\rangle \\
      |\nu_\mu (\vec{k}) \nu_e(\vec{p})\rangle \\
      |\nu_\mu (\vec{k}) \nu_\mu(\vec{p})\rangle
    \end{array}\right)
=\frac{\sqrt{2}G_F}{V} (1-\cos\Theta) \left(
\begin{array}{cccc}
  2 & 0 & 0 & 0 \\
  0 & 1 & 1 & 0 \\
  0 & 1 & 1 & 0 \\
  0 & 0 & 0 & 2
\end{array}
\right) \left(
    \begin{array}{c}
      |\nu_e (\vec{k}) \nu_e(\vec{p})\rangle  \\
      |\nu_e (\vec{k}) \nu_\mu(\vec{p})\rangle \\
      |\nu_\mu (\vec{k}) \nu_e(\vec{p})\rangle \\
      |\nu_\mu (\vec{k}) \nu_\mu(\vec{p})\rangle
    \end{array}\right),
\end{eqnarray}
where $V$ is the normalization volume. The value of the angular
factor, $(1-\cos\Theta)$, follows from (\ref{eq:gen_angle})
(taking into account the anticommutativity of the spinors), while
the flavor structure of the interaction is determined by
Eq.~(\ref{eq:spinH}).

The condition $\beta=\Theta$ by itself, however, \emph{does not
guarantee} coherence.
We can see that by considering the scattering of an electron
neutrino on two ``background" muon neutrinos.
Suppose that the initial momenta of the muon neutrinos are
orthogonal to that of the electron neutrino and, further, that the
neutrino wave packets are sufficiently narrow so that the two
scattering events are independent, with each given by
Eq.~(\ref{eq:pant4by4}). Then, as a result of the interaction,
we get
\begin{equation}\label{eq:2neutrino}
    |\nu_e\rangle |\nu_\mu\nu_\mu\rangle \longrightarrow
    |F\rangle=(1+2ia)|\nu_e\rangle |\nu_\mu\nu_\mu\rangle+
    i a(|\nu_\mu\rangle |\nu_e\nu_\mu\rangle +
    |\nu_\mu\rangle |\nu_\mu\nu_e\rangle),
\end{equation}
where $a\equiv -\sqrt{2}G_F(1-\cos\Theta)t/V$.

Both the interaction time $t$ and the volume $V$ depend on the
size of the wavepackets. Taking the wavepackets for simplicity to
be spheres of size $l$, we can write
\begin{eqnarray}
  t &\sim& l,
  V \sim l^3,\\
  a&\sim& -G_F(1-\cos\Theta)/l^2.
\end{eqnarray}
We notice that $|a| \ll 1$, as long as the size of the wavepackets
is much greater than (100 GeV)$^{-1}$, \textit{i.e.}, the
neutrino energy is well below the weak scale. This justifies
neglecting terms of order $a^2$ and higher in Eq.
(\ref{eq:2neutrino}).

Let us compute the probability $P$ that, as a result of the interaction,
the incident neutrino was converted into $\nu_\mu$. Introduce the
``$\nu_\mu$ number" operator $\hat \mu \equiv
|\mu\rangle\langle\mu|$ that acts only on the state of the
incident neutrino. Using the orthogonality of the states
$|\nu_e\nu_\mu\rangle$ and $|\nu_\mu\nu_e\rangle$, the probability
in question is
\begin{equation}\label{eq:P2neutrino}
    P=\langle F| \hat \mu|F\rangle = 2 a^2.
\end{equation}
This shows that the process is \emph{incoherent}: the probability
we found coincides with the sum of the conversion probabilities
for each scattering (a coherent process would yield $P=(2a)^2$).
As can be easily confirmed, as the number of ``background"
neutrinos $N$ is increased, the probability of flavor conversion
goes as $N a^2$.

It can be further shown that if the background neutrinos are in a
flavor superposition state
$\nu_x=\cos\alpha\nu_e+\sin\alpha\nu_\mu$, the probability of
conversion is \emph{not}, in general, equal to the sum of the
probabilities for each scattering. Indeed, a straightforward
calculation yields $P=a^2\sin^2\alpha(2+2\cos^2\alpha)$. Comparing
this with the conversion probability for a single scattering
event, $a^2\sin^2\alpha$, we see that the conversion process is
\emph{partially coherent},  and the degree of coherence increases
with decreasing $\alpha$. A generalization to a larger number of
scatterers is \cite{us}
\begin{equation}
  \label{eq:PN}
  P=a^2 \sin^2 \alpha ((N^2-N) \cos^2\alpha + N),
\end{equation}
which clearly shows the presence of partial coherence (the terms
proportional to $N^2$).

The above example shows that whether a neutrino undergoes coherent
conversion or not depends on the flavor state of the many background
neutrinos it interacts with. Moreover, as can be seen from
Eq.~(\ref{eq:2neutrino}), each neutrino-neutrino interaction creates
an entangled state.  Hence, the flavor evolution of a system of
several interacting neutrinos demands a many-particle description.
The question is: does the wavefunction of the system somehow factorize
into individual one-particle wavefunctions in the limit of large $N$,
and, if yes, under what conditions?  Below we describe two recently
proposed approaches to this problem.

\section{Single- vs. many-particle description: two approaches}
\label{sect:twoapproaches}

\subsection{First approach: interference of many elementary scattering amplitudes}
\label{subsect:singlevsmany}

The first method was developed in \cite{us}. The idea is to
describe the flavor evolution in a neutrino gas by generalizing
the procedure of adding elementary scattering amplitudes, as was
outlined in the three-neutrino example of Sect.
\ref{subsect:cohvsincoh}.

Let us consider the interactions between two orthogonal neutrino
beams. Once the results for this setup are understood, they can be
straightforwardly generalized to more complicated systems. Let us
assume that the neutrino wavepacket size is much smaller than the
particle spacing ($l \ll n^{-1/3}_\nu$), so that different
neutrino-neutrino interactions can be treated independently using
Eq.~(\ref{eq:pant4by4}). As an initial configuration, we take the
first beam to be made of $N_1$ electron neutrinos $\nu_e$, and the
second beam to be made of $N_2$ neutrinos in the flavor
superposition state
$\nu_x\equiv\cos\alpha\nu_e+\sin\alpha\nu_\mu$. It can be shown
\cite{us} that the result of the interaction during a small time
$\delta t$ can be written as
\begin{eqnarray}
\label{eq:2b_exch}
|e e e ...\rangle
  |x x x ...\rangle
&\stackrel{t\rightarrow t+\delta t}{\longrightarrow}& |F\rangle=
  |e e e ...\rangle
  |x x x ...\rangle
+ i a |F_1\rangle,\\
  \label{eq:2beams2_full}
 |F_1\rangle &=& N_1 N_2(1+ |\langle e |x \rangle|^2 ) |e e e
...\rangle
|x x x ...\rangle \nonumber\\
&+& N_2 \langle\mu |x \rangle \langle x |e \rangle (|\mu e e
...\rangle+ |e \mu e ...\rangle+...)
|x x x ...\rangle\nonumber\\
&+& N_1 \langle e |x \rangle \langle y  |e \rangle |e e e
...\rangle (|y  x x ...\rangle+ |x y  x ...\rangle+...)
\nonumber\\
&+& \langle\mu |x \rangle \langle y  |e \rangle
 (|\mu e e ...\rangle+|e \mu e ...\rangle+...)
 (|y  x x ...\rangle+|x y  x ...\rangle+...),\;\;\;\;\;
\end{eqnarray}
a direct generalization of Eq.~(\ref{eq:2neutrino}).
Here the state $|y \rangle$ is defined to be orthogonal to $|x \rangle$:
 $\langle x | y \rangle = 0$.

The evolution in Eqs.~(\ref{eq:2b_exch},\ref{eq:2beams2_full})
cannot be described using only one-particle equations, since the
final state in Eq.~(\ref{eq:2beams2_full}) is an entangled
many-particle state. Nevertheless, as observed in \cite{us},
\emph{the coherent part} of the evolution can be. Indeed, the last
term in Eq.~(\ref{eq:2beams2_full}) represents an incoherent
effect, since it contains the sums of mutually orthogonal terms,
and could be dropped. The sum of the  remaining three terms  is
equal, to the first order in $a$, to a product of rotated
single-particle states.

Since for small $\delta t$ a state of each neutrino undergoes a
small rotation and since nothing in this argument depends on the
particular choice of the initial states,
Eqs.~(\ref{eq:2b_exch},\ref{eq:2beams2_full}) specify how the
states will rotate at any point in time. This observation means
that, for each neutrino, one can write a differential equation
describing its evolution for any --- not necessarily small ---
$t$. This equation takes on the form \cite{us}
\begin{eqnarray}
\label{eq:oureqn}
 i \frac{d|\psi^{(i)}\rangle}{d t}&=&H_{aa}|\psi^{(i)}\rangle,\nonumber\\
H_{aa}&=&\sum_{j} \sqrt{2} G_F
n^{(j)}(1-\cos\Theta^{(ij)})\left[|\phi^{(j)}\rangle\langle\phi^{(j)}|
+\frac{1}{2}-\frac{1}{2} |\langle\phi^{(j)}|\psi^{(i)}\rangle|^2 \right],
\end{eqnarray}
where, for generality, we restored the angular factor and summed over
possible angles $\Theta^{(ij)}$. The flavor-space wavefunction of the
given neutrino is denoted by $\psi^{(i)}$, those of the other
(``background") neutrinos by $\phi^{(j)}$. The quantity $n^{(j)}$
denotes the number density of neutrinos whose momenta make an angle
$\Theta^{(ij)}$ with the given neutrino.  This equation is similar to
the one-particle evolution equations given in
\cite{Pantaleone1992prd,germanpaper,McKellarThomson}. In fact, the
only difference is in the last two terms in the square brackets,
which have no effect on flavor evolution.

Let us discuss 
a particular case
when the neutrinos at $t=0$ are in the flavor eigenstates. We note
that, according to Eq.~(\ref{eq:oureqn}), a coherent conversion
effect in this case is absent. This result in turn could be traced
to Eq.~(\ref{eq:2beams2_full}), where the second and the third
terms  vanish for $\nu_x=\nu_\mu$ or $\nu_x=\nu_e$. The only
remaining source of flavor transformation is the last term in
Eq.~(\ref{eq:2beams2_full}), which is nonzero for $\nu_x=\nu_\mu$.
Since this term represents incoherent scattering, for a neutrino
traveling through a gas of neutrinos of opposite flavor the
conversion probability for small $t$ has the form $P(t)=N(t) a^2$.

At first, it is not obvious how this incoherent effect is related
to the incoherent flavor exchange that occurs in a real neutrino
ensemble.  Indeed, as mentioned in the introduction to
Eq.~(\ref{eq:pant4by4}), by replacing the true interaction
Eq.~(\ref{eq:NChamiltonian}) with the forward scattering
Hamiltonian, we have truncated most of the incoherent effects.
Nevertheless, a connection between the two can be made.  The
argument, which follows below, is very instructive and will prove
helpful in the subsequent Sections.

When discussing the ``forward scattering probability", we in fact mean
the probability for the neutrino to be scattered in a certain (small)
solid angle $\delta \Omega$ around the direction of the incoming
neutrino (the probability of neutrino scattering at any fixed angle
is, of course, zero). A neutrino wave packet with the cross section
area $A$ and wavelength $\lambda=2\pi/E_\nu$ is diverging due to
diffraction into a solid angle $\delta\Omega_{\rm diff}=\lambda^2/A$.
The neutrino wave which is scattered into the same solid angle $\delta
\Omega$ will interfere with the incident wave. Thus, $\delta
\Omega=\delta \Omega_{\rm diff}$.

With this in mind, let us consider the result of forward
scattering of a given neutrino on $N$ other neutrinos in the
medium. According to Eq.~(\ref{eq:pant4by4}), for neutrino wave
packets of longitudinal size $l$ and cross section $A$, $a\sim -G_F
l(1-\cos\Theta)/Al=-G_F(1-\cos\Theta)/A$. The resulting flavor
conversion probability $P$ depends on whether the forward
scattering amplitudes add up coherently or incoherently. In the
first case, $P_{\rm coh}\sim N^2 a^2$, in the second, $P_{\rm
inc}\sim N a^2$. The number of interactions during time $t$ is
$N=A t n_\nu$ ($n_\nu$ is the neutrino number density) and we find
that in the case of coherent forward scattering the dependence on
the size of the wave packet \emph{drops out},
\begin{equation}\label{eq:Pcoh}
    P_{\rm coh}\sim \frac{G_F^2(1-\cos\Theta)^2}{A^2} A^2 t^2 n_\nu^2
    =G_F^2(1-\cos\Theta)^2 t^2 n_\nu^2.
\end{equation}
This is indeed seen from Eq.~(\ref{eq:oureqn}), which does not
contain the size of the neutrino wavepacket.

On the other hand, for the case of incoherent forward scattering,
the cross section of the wave packet does enter the final result,
\begin{equation}\label{eq:Pinc}
    P_{\rm inc}\sim \frac{G_F^2(1-\cos\Theta)^2}{A^2} A t n_\nu
    =\frac{G_F^2(1-\cos\Theta)^2}{A} t n_\nu.
\end{equation}
The dependence on the size of the wavepacket look puzzling.
Nevertheless, it is easily understood once we recall that, by
construction, the result in (\ref{eq:Pinc}) represents the
fraction of incoherent scattering events for which one of the two
neutrinos is scattered in the forward cone
$\delta\Omega\sim\lambda^2/A\propto 1/E_\nu^2 A$. This means that,
up to a numerical coefficient, the probability of incoherent
scattering in any direction $P_{\rm inc}^{\rm TOT}$ is given by
\begin{equation}\label{eq:PTOT}
  P_{\rm inc}^{\rm TOT} \sim P_{\rm inc}/\delta\Omega
  \sim G_F^2 E_\nu^2(1-\cos\Theta)^2 t n_\nu.
\end{equation}
This has the form $\sigma_w t n_\nu$, where $\sigma_w$ is a
typical weak interaction cross section, $\sigma_w \propto G_F^2
E_\nu^2$, exactly as one would expect.

Thus, in summary, for the purpose of studying coherent effects, the
full Hamiltonian (\ref{eq:NChamiltonian}) may be replaced by the
forward scattering flavor exchange Hamiltonian, as done in
Eq.~(\ref{eq:pant4by4}), with the understanding that in the final
result the terms that go like $N^2$ represent the \emph{complete}
coherent effect while the terms that go like $N$ correspond to only the
\emph{fraction} of the incoherent scattering
that occurs in the forward diffraction cone.

\subsection{Second approach: solving a many-particle equation for neutrino plane waves in a box}
\label{subsect:reviewbell}

An alternative method of studying coherent processes in a neutrino
gas was proposed in \cite{Belletal}. The idea is to consider a
system of interacting neutrino plane waves in a box and study its
flavor evolution, again with the restriction to forward scattering
only. The neutrinos are taken to be initially in flavor states and
the goal is to test the existence of coherent conversion effects
by looking at the rate of equilibration of the system.

The ``forward scattering" Hamiltonian can be found by generalizing
Eq.~(\ref{eq:pant4by4}).  A basis of states is formed by the
initial configuration of neutrinos and all its possible
permutations. Taking, e.g., $N$ electron and $N$ muon neutrinos,
there are $(2N)!/(N!)^2$ distinct basis states. The entries of the
Hamiltonian are found as follows. (i) The diagonal entries receive
$N(3N-2)$ contributions (this includes $(2N)!/[2!(2N-2)!]=N(2N-1)$
flavor-blind scattering processes with scattering angle $\beta=0$
and $2N!/[2!(N-2)!]=N(N-1)$ processes with scattering angle
$\beta=\Theta$). (ii) Each off-diagonal entry receives a single
contribution from the exchange process that connects the two
corresponding basis states; if the two basis states are not
connected by a single permutation, the corresponding off-diagonal
entry vanishes. (iii) Each contribution equals $\sqrt{2} G_F
(1-\cos\Theta_{ij})/V$, where $\Theta_{ij}$ is the angle between
the momenta of the two interacting neutrinos.

For example, for the case of $N=2$ there are $6$ basis states:
$|\nu_e\nu_e\nu_\mu\nu_\mu\rangle$,
$|\nu_\mu\nu_e\nu_e\nu_\mu\rangle$,
$|\nu_\mu\nu_e\nu_\mu\nu_e\rangle$,
$|\nu_e\nu_\mu\nu_e\nu_\mu\rangle$,
$|\nu_e\nu_\mu\nu_\mu\nu_e\rangle$,
$|\nu_\mu\nu_\mu\nu_e\nu_e\rangle$, and the Hamiltonian is given
by
\begin{eqnarray}
H_{N=2}=\frac{\sqrt{2} G_F}{V}\left(
\begin{array}{llllll}
    d_{12,34}&f_{13}   &f_{14}   &f_{23}   &f_{24}   &0 \\
    f_{13}   &d_{14,23}&f_{34}   &f_{12}   &0        &f_{24} \\
    f_{14}   &f_{34}   &d_{13,24}&0        &f_{12}   &f_{23} \\
    f_{23}   &f_{12}   &0        &d_{13,24}&f_{34}   &f_{14} \\
    f_{24}   &0        &f_{12}   &f_{34}   &d_{14,23}&f_{13} \\
    0        &f_{24}   &f_{23}   &f_{14}   &f_{13}   &d_{12,34}
\end{array}\right).
\end{eqnarray}
Here $f_{ij}=1-\cos\Theta_{ij}$ is the angular factor for the
$i,j$ pair and $d_{ij,kl}=f_{ij}+f_{kl}+\sum$, where $\sum$ is the
sum of all distinct $f_{ij}$.
\\

Let us discuss the meaning of the equilibration time $t_{\rm eq}$.
If conversion is coherent, $t_{\rm eq}$ is defined as the time
over which a neutrino of a given momentum is expected to change
its flavor. The momentum itself stays unchanged over this time
scale. If conversion is incoherent, $t_{\rm eq}$ is defined as the
time over which the neutrino momenta will be randomized (each
neutrino can no longer be identified by its momentum).
In the model we are considering, however, the definition
of equilibration for coherent process  also applies to the incoherent
process, since the interaction is chosen to preserve the neutrino
momentum.

Let us determine what dependence of $t_{\rm eq}$ would be a sign
of coherent or incoherent conversion. To do this, it is useful to
repeat the exercise of Sect. \ref{subsect:singlevsmany} of
estimating the conversion probability as a function of time. For
neutrino plane waves in the box the following two modifications to
the argument need to be made: first, the normalization volume $V$
becomes the size of the box and, second, the interaction time
becomes the actual time $t$ elapsed from the beginning of the
evolution.  The scattering amplitude becomes $a\sim -G_F
t(1-\cos\Theta)/V$ and, since each neutrino simultaneously
interact with all the others, the number of interactions is $N=V
n_\nu$. Using these ingredients, we find
\begin{eqnarray}
  \label{eq:Pcoh_box}
    P_{\rm coh}&\sim& \frac{G_F^2(1-\cos\Theta)^2 t^2}{V^2} V^2 n_\nu^2
    =G_F^2(1-\cos\Theta)^2 t^2 n_\nu^2,\\
  \label{eq:Pinc_box}
    P_{\rm inc}&\sim& \frac{G_F^2(1-\cos\Theta)^2 t^2}{V^2} V n_\nu
    =\frac{G_F^2(1-\cos\Theta)^2}{V} t^2 n_\nu.
\end{eqnarray}

Here, as in the case of the interacting neutrino wavepackets, we
find that in $ P_{\rm coh}$ the dependence on the size of
the box cancels out (and the result is the same as
Eq.~(\ref{eq:Pcoh})), while for incoherent scattering the size of
the box enters the final result. Since, as was already stressed,
the incoherent scattering result is, in some sense, artificial,
there is no cause for alarm.

Inverting Eqs.~(\ref{eq:Pcoh_box},\ref{eq:Pinc_box}) and
Eqs.~(\ref{eq:Pcoh},\ref{eq:Pinc}), we find that the equilibration
time for coherent scattering in both cases has the dependence
\begin{equation}\label{eq:tcoh_form}
    t_{\rm eq}^{coh}=(G_F n_\nu)^{-1},
\end{equation}
while for incoherent scattering one finds
\begin{equation}\label{eq:t_inc_WP}
    t_{\rm eq}^{inc}=(G_F^2 n_\nu/A)^{-1}
\end{equation}
for the case of neutrino
wavepackets,
and
\begin{equation}\label{eq:t_inc_BOX}
    t_{\rm eq}^{inc}=(G_F \sqrt{n_\nu/V})^{-1}
\end{equation}
for plane waves in a box.
In what follows, we solve the evolution of the system of neutrinos
in the box completely analytically.
This gives the equilibration time $t_{eq}$, which  we compare  to
Eqs.~(\ref{eq:tcoh_form}) and (\ref{eq:t_inc_BOX}) to determine
the character of the evolution.

\section{Plane waves in a box: analytical treatment}
\label{sec:full}

\subsection{Setup}
\label{sec:setup}

As done in \cite{Belletal}, we consider a system of $2N$ neutrino
plane waves, of which $N$ are initially in the $\nu_e$ state and
the remaining $N$ are in the $\nu_\mu$ state.  We write this
configuration of the system as:
\begin{equation}\label{eq:initstate_nu}
    |S(0)\rangle=|\underbrace{\nu_e \nu_e...\nu_e}_N
    \underbrace{\nu_\mu\nu_\mu...\nu_\mu}_N\rangle~.
\end{equation}
Here each neutrino wave is identified by its position in the list,
which corresponds to a given momentum.

The Hamiltonian for this system has the form described in
Sect.~\ref{subsect:reviewbell}.  Since it is equivalent to the
Hamiltonian of a system of spins (Sect. \ref{subsect:flavor}), in
the following we adopt the terminology of angular momenta. As a
convention, we identify the ``up" (pointing along the $z$ axis)
state of a spin with $\nu_e$ and the ``down" state with $\nu_\mu$.
Thus, the initial state (\ref{eq:initstate_nu}) can be expressed
as
\begin{equation}\label{eq:initstate}
    |S(0)\rangle=|\underbrace{+\frac{1}{2},+\frac{1}{2},...+\frac{1}{2}}_N,
    \underbrace{-\frac{1}{2},-\frac{1}{2},...-\frac{1}{2}}_N\rangle.
\end{equation}
To make an analytical treatment possible, we set \be f_{ij}=1
\label{fij} \ee for all neutrino pairs. The resulting Hamiltonian
describes an ensemble of spins interacting with each other with
equal strength\footnote{This makes it different from the familiar
Ising model.}.  This simplifying assumption does not change the
coherent (or incoherent) character of the conversion effects, as
also pointed out in \cite{Belletal}. With this assumption, the
nonzero off-diagonal entries of the Hamiltonian become equal to
$\sqrt{2}G_F/V$ while the diagonal entries become
$\sqrt{2}G_F/V\times N(3N-2)$.

We recall (Sect. \ref{subsect:singlevsmany}) that for the initial
configuration (\ref{eq:initstate_nu}) the single-particle
formalism predicts no coherent effects, as can be seen from Eq.
(\ref{eq:oureqn}). We investigate if this conclusion is changed
once the simultaneous interactions between all neutrinos are
considered.

\subsection{Results}
\label{sec:res}

Let us first display our results.  Their derivation and a more
detailed discussion is given in Sec. \ref{sec:der}.

The probability $P_1$ that the first spin -- which is in the state
$|+1/2\rangle$ at $t=0$ -- is found in the same up state at time
$t$ ($t>0$) can be calculated by a completely analytical
procedure. It is given by the Fourier series:
\begin{eqnarray}
\label{eq:main_answer} P_1(t)=\frac{1}{2}+\frac{1}{2}
\sum_{J=0}^{N-1}
    \eta(N,J)\cos[(2J+2)g t]~,
\end{eqnarray}
where $g\equiv \sqrt{2} G_F/V$.
The coefficients $\eta(N,J)$ are the squares of the Clebsch-Gordan
coefficients $\langle j_1 m_1 j_2 m_2| J M\rangle\equiv C(j_1,
m_1, j_2, m_2, J, M)$ for the addition of two angular momenta,
$j_1=m_1=(N-1)/2$ and $j_2=-m_2=N/2$:
\begin{eqnarray}
\label{eq:coeff}
 \eta(N,J) &=& \left[C\left(\frac{N-1}{2}, \frac{N-1}{2},
 \frac{N}{2},-\frac{N}{2}, J+\frac{1}{2},-\frac{1}{2}\right)\right]^2 \nonumber\\
 &=&
 \frac{(1 + 2 J) N [(N-1)!]^2}{(J + N+1)!(N-J-1)!}.
\end{eqnarray}
We follow the standard notation, in which $j_1,j_2$ denote the
values of the angular momenta being added, $m_1,m_2$ denote their
projections along the $\hat z$ direction, and $J$ and $M$ are the
corresponding values for the total spin.

\begin{figure}
\includegraphics[width=0.9\textwidth]{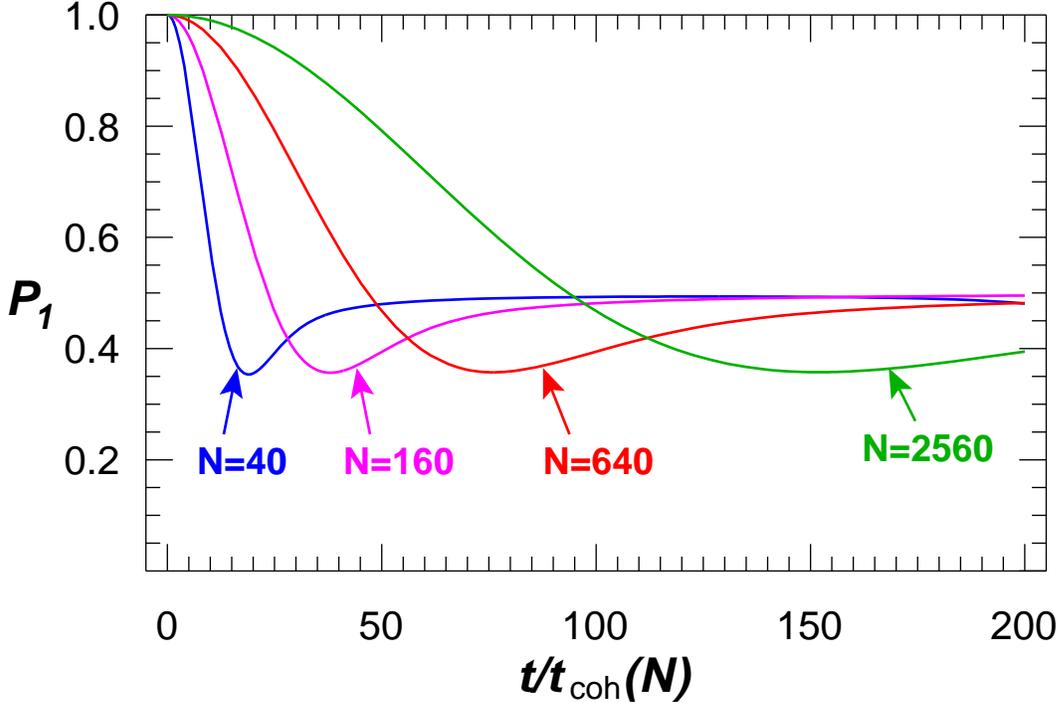}\\
\caption{The time-evolution of the survival probability $P_1$,
for different values of $N$ (numbers on the curves).  The unit of
the horizontal axis is chosen to be $t/t^{coh}_{eq}=2 t g N$, as
used in \cite{Belletal}. If the conversion were coherent,  the
curves should reach equilibration at the same point on this scale.
} \label{fig:largeNplot}
\end{figure}
Figure \ref{fig:largeNplot} shows the survival probability
$P_1(t)$, obtained by summing numerically the series
(\ref{eq:main_answer}). The horizontal axis represents time in
units of the equilibration time predicted for coherent conversion,
$t^{coh}_{eq} = (2N  g )^{-1}$ (Eq. (\ref{eq:tcoh_form}) for $2N$
neutrinos).
The curves refer to $N=40, 160, 640, 2560$. Having the same choice
of units, our plot can be directly compared to fig. 2 of ref.
\cite{Belletal}, where, however, smaller values of $N$ were
used\footnote{In ref. \cite{Belletal} a different form of the
spin-spin (neutrino-neutrino) coupling was adopted, namely, the
angular factors were randomly generated. This produces a numerical
difference with respect to our results, but does not affect the
conclusions about the equilibration time.}. From Fig.
\ref{fig:largeNplot}, it is clear that the equilibration time
$t_{eq}$ is larger than $t^{coh}_{eq}$ and the discrepancy grows
with $N$.

In the limit $N\gg 1$  the series (\ref{eq:main_answer}) is well
approximated by the following function:
\begin{eqnarray}\label{eq:final_beauty}
    P_1(t)=1-\frac{\sqrt{\pi}}{2}\sqrt{N}g t\exp(-N g^2 t^2)
    {\rm erfi}(\sqrt{N}g t),
\end{eqnarray}
where ${\rm erfi}(z)$ is the imaginary error function
\begin{equation}
{\rm erfi}(z)\equiv -i~{\rm erf}(iz)=
\frac{2}{\sqrt{\pi}}\times\int_0^z \exp(t^2)dt~. \label{erfi}
\end{equation}
We have checked that for the values of $N$ shown in
Fig.~\ref{fig:largeNplot} the approximation is very good and the
curves obtained using (\ref{eq:final_beauty}) are
indistinguishable from those obtained from Eq.
(\ref{eq:main_answer}).

From Eq. (\ref{eq:final_beauty}) we see that the characteristic
equilibration time equals:
\begin{equation}
t_{eq} = (g\sqrt{N})^{-1}~. \label{teq}
\end{equation}
This scaling of the equilibration time can be clearly seen in
Fig.~\ref{fig:largeNplot}. It is precisely (up to numerical
factors) the form (\ref{eq:t_inc_BOX}) expected if the
equilibration process is incoherent.

\subsection{Derivation of results}
\label{sec:der}

\subsubsection{The eigenvalues of the Hamiltonian}

From Eqs. (\ref{eq:spinH}) and (\ref{fij}) it follows that for
each pair $i$ and $j$ the interaction is given by
$H_{ij}=g\hat{L}_{ij}^2=g(\vec\sigma_i+\vec\sigma_j)^2$ and,
therefore, the Hamiltonian for the whole system is
\begin{equation}\label{eq:Hsystem}
    H_{N}=g\sum_{i=1}^{2N-1} \sum_{j=i+1}^{2N} (\vec\sigma_i+\vec\sigma_j)^2=
    g\left(\sum_{i=1}^{2N-1} \sum_{j=i+1}^{2N} 2\vec\sigma_i\vec\sigma_j+
    \frac{3}{2}N(2N-1)\right),
\end{equation}
where the summation is done over all $N(2N-1)$ pairs.

We notice that this Hamiltonian can be related to the operator of
the square of the total angular momentum of the system,
\begin{equation}\label{eq:Ltot}
    \hat{L}^2\equiv\left(\sum_{i=1}^{2N} \vec\sigma_i\right)^2=
    \sum_{i=1}^{2N-1} \sum_{j=i+1}^{2N} 2\vec\sigma_i\vec\sigma_j+\frac{3}{2}N.
\end{equation}
By comparing Eqs.~(\ref{eq:Hsystem}) and (\ref{eq:Ltot}) we find
\begin{equation}\label{eq:HsystemLtot}
    H_{N}=g [\hat{L}^2+3N(N-1)],
\end{equation}
which has the eigenvalues
\begin{equation}\label{eq:Esystem}
    E_{N,J}=g [J(J+1)+3N(N-1)], \hskip0.5truecm {\rm with} \hskip0.5truecm
J=0,1,...,N~~.
\end{equation}
Since the flavor evolution will depend only on the difference of
the eigenvalues (as will be shown later), in the following we will
omit the second term in Eq.~(\ref{eq:Esystem}) and write
\begin{equation}
E_J = g J(J+1)~ . \label{eig}
\end{equation}

A given value of $E_J$ in general corresponds to more than one
state. This degeneracy can be seen by simply counting the states.
For our specific setup, with $M=0$, there are $(2N)!/(N!)^2$ basis
states, while there are only $N+1$ different values of $J$.  In
general, therefore, an orthonormal basis of eigenstates of the
Hamiltonian has the form:
\begin{equation}\label{eq:fullbasis}
    |J, k_J\rangle,
\end{equation}
where the index $k_J$ labels eigenstates with the same value of
$J$.

As an illustration of the present discussion, we give the
Hamiltonian for the case $N=2$ of Sect. \ref{subsect:reviewbell}:
\begin{eqnarray}
H_{N=2}=g\left(
\begin{array}{llllll}
    8 & 1 & 1 & 1 & 1 &0 \\
    1 & 8 & 1 & 1 & 0 &1 \\
    1 & 1 & 8 & 0 & 1 &1 \\
    1 & 1 & 0 & 8 & 1 &1 \\
    1 & 0 & 1 & 1 & 8 &1 \\
    0 & 1 & 1 & 1 & 1 &8
\end{array}\right).
\end{eqnarray}
The eigenvalues, which are degenerate as expected, are
$g(6,6,8,8,8,12)$. They correspond to the following values of $J$:
$(0,0,1,1,1,2)$, as can be seen from Eq.~(\ref{eq:Esystem}).

\subsubsection{The Fourier series}
\label{sec:four} 
Let us start with proving Eq. (\ref{eq:main_answer}).

In the interest of clarity, we first display some definitions.
Let us denote as $|+1/2\rangle_1$ and $|-1/2\rangle_1$ the ``up''
and ``down'' states of the first spin. Given a state $|\psi
\rangle$ of the $2N$ spins of the system, it will be convenient to
use the projections $\langle +1/2|_1|\psi \rangle$ and $\langle
-1/2|_1|\psi \rangle$, which are states of $2N-1$ spins.
In particular, we introduce the state $|S^-\rangle$ obtained by
removing the first spin from the initial configuration $| S(0)
\rangle$ (Eq. (\ref{eq:initstate})):
\begin{eqnarray}
\langle +1/2|_1|S(0) \rangle \equiv |S^-\rangle \equiv
  |\underbrace{+\frac{1}{2},+\frac{1}{2},...+\frac{1}{2}}_{N-1},
  \underbrace{-\frac{1}{2},-\frac{1}{2},...-\frac{1}{2}}_N\rangle~.
\label{smen}
\end{eqnarray}

For any state, we can define another state obtained from the first
one upon a reflection along the $z$-axis. In this transformation, each
constituent spin $| +1/2 \rangle $ is transformed into spin $|
-1/2 \rangle $ and vice versa.
For instance, the application of this transformation to $|S^-\rangle$
gives its ``mirror'' state:
\begin{eqnarray}
 | S^+ \rangle\equiv
|\underbrace{-\frac{1}{2},-\frac{1}{2},...-\frac{1}{2}}_{N-1}
  \underbrace{+\frac{1}{2},+\frac{1}{2},...+\frac{1}{2}}_{N}\rangle~.
\label{splus}
\end{eqnarray}
In what follows we will exploit the properties of various states under
this reflection.

We adopt the notation $|j,m,s\rangle_{n}$ to indicate the result
of the following procedure: consider the state $| s \rangle$ and
take its projection on the subspace formed by the states of $n$
spins that have the total angular momentum $j$ and projection $m$
along the $\hat z$ axis; normalize the result to 1. As an example, $|
J, 0, S(0)\rangle_{2N}$ indicates the normalized projection of
$S|(0)\rangle$ on the subspace of the states of $2N$ spins having
total angular momentum $J$ and $\hat z$-projection equal to zero.

It is also convenient to define the state:
\begin{equation}
|s^-_J\rangle \equiv \langle +1/2|_1| J, 0, S(0)\rangle_{2N}~,
\label{sj}
\end{equation}
(which has $2N-1$ spins) and denote as $|s^+_J\rangle$ its mirror
state obtained upon $\hat{z}$-reflection.
\\

The proof is constituted by four parts.\\

\noindent 1. \underline{Expansion of the initial state}\\

The initial state can be expanded on the basis of the eigenstates
of the Hamiltonian:
\begin{equation}\label{eq:S0}
    |S(0)\rangle = \sum_{J,k_J} | J, k_J\rangle\langle J,
    k_J|S(0)\rangle,
\end{equation}
which then can be evolved in time as
\begin{equation}\label{eq:ev_S_eig}
    |S(t)\rangle = \sum_{J,k_J} e^{-i E_J t}| J, k_J\rangle\langle J, k_J|S(0)\rangle.
\end{equation}
Since the energy depends only on $J$, 
the summation over $k_J$ can be done, yielding
\begin{eqnarray}
&& |S(0)\rangle = \sum_{J=0}^N A_J | J, 0, S(0)\rangle_{2N}~, \nonumber \\
&& |S(t)\rangle = \sum_{J=0}^N A_J e^{-i E_J t}| J, 0,
S(0)\rangle_{2N}~. \label{eq:ev_S_in}
\end{eqnarray}
The factors $A_J$ equal the Clebsch-Gordan coefficients:
\begin{equation}\label{eq:Aj_C}
    A_J=C\left(\frac{N}{2}, \frac{N}{2},
 \frac{N}{2},-\frac{N}{2}, J,0\right).
\end{equation}
This can be understood considering that the initial state,
$|S(0)\rangle$, equals the sum of two angular momenta with
$j_1=N/2$, $j_2=N/2$ and projections $m_1=N/2$, $m_2=-N/2$. The
first momentum is given by the sum of all the  ``up'' spins, while
the second is the sum of all the ``down'' spins.
\\

\noindent 2.
\underline{Decomposition into states of $1$ and $2N-1$ spins} \\

It is useful to decompose the state $|S^-\rangle$ as:
\begin{equation}\label{eq:S0_1+2N-1}
    |S^-\rangle=
    \sum_{J=0}^{N-1}
    K_{J+1/2}|J+\frac{1}{2},-\frac{1}{2}, S^-\rangle_{2N-1}~.
\end{equation}
Similarly to Eq. (\ref{eq:Aj_C}), here we have:
\begin{equation}\label{eq:1st_relation}
    K_{J+1/2}=C\left(\frac{N-1}{2}, \frac{N-1}{2},
 \frac{N}{2},-\frac{N}{2}, J+\frac{1}{2},-\frac{1}{2}\right)=\sqrt{\eta(N,J)}~,
\end{equation}
where $\eta(N,J)$ was given in Eq. (\ref{eq:coeff}).  The argument
to explain (\ref{eq:1st_relation}) is analogous to that used to
justify Eq.  (\ref{eq:Aj_C}): since the state $|S^-\rangle$ has
$N-1$ spins ``up'' and N spins ``down'', it is given by the sum of
two angular momenta with $j_1=(N-1)/2$, $j_2=N/2$ and projections
$m_1=(N-1)/2$, $m_2=-N/2$. These momenta are the result of summing
all the spins ``up" and all the spins ``down" separately.

Each of the states $| J, 0,
S(0)\rangle_{2N}$ can be decomposed as:
\begin{eqnarray}
\label{eq:decomp1}
    |J,0,S(0)\rangle_{2N} &=& a_J|+\frac{1}{2}\rangle_1
    |J+\frac{1}{2},-\frac{1}{2},s^-_J\rangle_{2N-1}+
    b_J|+\frac{1}{2}\rangle_1
    |J-\frac{1}{2},-\frac{1}{2},s^-_J\rangle_{2N-1}\nonumber\\ &+&
    c_J|-\frac{1}{2}\rangle_1
    |J+\frac{1}{2},+\frac{1}{2},s^+_J\rangle_{2N-1}+
    d_J|-\frac{1}{2}\rangle_1
    |J-\frac{1}{2},+\frac{1}{2},s^+_J\rangle_{2N-1}~,~~~
\end{eqnarray}
where the states $ |J+\frac{1}{2},+\frac{1}{2},s^+_J\rangle_{2N-1}$
and $ |J-\frac{1}{2},+\frac{1}{2},s^+_J\rangle_{2N-1}$ in the second
line are related to the corresponding ones in the first line by the
$z$-reflection operation.
Notice that the state $|J,0,S(0)\rangle_{2N}$ has parity $(-1)^{N-J}$
under $z$-reflection. Requiring that the right hand side of
Eq. (\ref{eq:decomp1}) has the same parity yields
\begin{eqnarray}
 &&c_J=(-1)^{N-J} a_J \hskip1truecm d_J=(-1)^{N-J} b_J ~.
\label{oss1}
\end{eqnarray}
We also note that
\begin{eqnarray}
&&a_{N}=c_{N}=0 \hskip1truecm b_{0}=d_{0}=0~,
\label{oss2} \\
&&|a_J|^2+|b_J|^2=|c_J|^2+|d_J|^2=1/2~, \label{oss3}
\end{eqnarray}
where the last relation represents the normalization
condition.
\\

\noindent
3. \underline{Properties of the decomposition (\ref{eq:decomp1}).}\\

We can calculate the state $\langle-1/2|_1|S(t)\rangle$ and demand
that it vanishes at $t=0$.
This is equivalent to requiring a null probability to find the first
spin initially in the down state. Combining this condition
with Eqs. (\ref{eq:ev_S_in}), (\ref{eq:decomp1}) and (\ref{oss1})
we find
\begin{eqnarray}
   && \sum_{J=0}^{N-1} A_Ja_J(-1)^{N-J} |J+\frac{1}{2},+\frac{1}{2},s^+_J\rangle_{2N-1} \nonumber\\
  &&+  \sum_{J=0}^{N-1} A_{J+1}b_{J+1}(-1)^{N-J-1}
  |J+\frac{1}{2},+\frac{1}{2} ,s^+_{J+1}\rangle_{2N-1} =0~,
\label{eq:sum2}
\end{eqnarray}
from which it follows that 
\be
A_Ja_J|J+\frac{1}{2},+\frac{1}{2},s^+_J\rangle_{2N-1}
=A_{J+1}b_{J+1}|J+\frac{1}{2},+\frac{1}{2} ,s^+_{J+1}\rangle_{2N-1}~,
\label{aeqb} 
\ee 
and, by $z$-reflection: 
\be
A_Ja_J|J+\frac{1}{2},-\frac{1}{2},s^-_J\rangle_{2N-1}
=A_{J+1}b_{J+1}|J+\frac{1}{2},-\frac{1}{2} ,s^-_{J+1}\rangle_{2N-1}~.
\label{aeqb_refl} 
\ee

\noindent 4. \underline{Calculation of the survival probability 
$P_1$.}\\

Let us calculate the probability $P_1(t)$ that at the time $t$ the
first spin is found in  its initial state, $| +1/2 \rangle$:
\begin{equation}\label{eq:def_P1}
    P_1(t)= \langle S(t)|+1/2\rangle_1\langle+1/2|_1|S(t)\rangle~.
\end{equation}
The combination of Eqs. (\ref{eq:ev_S_in}), (\ref{eq:decomp1}) and
(\ref{oss2}), gives the expression of $
\langle+1/2|_1|S(t)\rangle$:
\begin{eqnarray}\label{eq:sum1}
 \langle+1/2|_1|S(t)\rangle
=&&  \sum_{J=0}^{N-1} e^{-i E_J t} A_Ja_J
    |J+\frac{1}{2},-\frac{1}{2}, s^-_J\rangle_{2N-1} \nonumber \\
&&+\sum_{J=0}^{N-1} e^{-i E_{J+1} t} A_{J+1}b_{J+1}
|J+\frac{1}{2},-\frac{1}{2},  s^-_{J+1}\rangle_{2N-1}~,
\end{eqnarray}
which, taking into account the result (\ref{aeqb_refl}), simplifies to:
\begin{eqnarray}\label{eq:sum1_sum}
 \langle+1/2|_1|S(t)\rangle
=  \sum_{J=0}^{N-1} \left(e^{-i E_J t} + e^{-i E_{J+1} t} \right)
A_Ja_J
     |J+\frac{1}{2},-\frac{1}{2}, s^-_J\rangle_{2N-1}~.
\end{eqnarray}

The final step to conclude the proof is the observation that for
$t=0$, Eq. (\ref{eq:sum1_sum}) should coincide with the expansion
of the state $| S^-\rangle$, given in Eq. (\ref{eq:S0_1+2N-1}).
From comparison, using the equality (\ref{eq:1st_relation}), it
follows that: 
\beq 
2 A_Ja_J |J+\frac{1}{2},-\frac{1}{2},
s^-_J\rangle_{2N-1}
&=&
K_{J+1/2} |J+\frac{1}{2},-\frac{1}{2},
S^-\rangle_{2N-1} \nonumber \\
&=&
\sqrt{\eta(N,J)} |J+\frac{1}{2},-\frac{1}{2},
S^-\rangle_{2N-1}~.
 \label{aeqs} 
\eeq

Combining Eqs. (\ref{eq:def_P1}), (\ref{eq:sum1_sum}), (\ref{aeqs}) and
(\ref{eig}), we find
\begin{equation}
  \label{eq:prelim_answer}
  P_1(t)=\sum_{J=0}^{N-1}\frac{\eta(N,J)^2}{4}
  |e^{-i E_J t} + e^{-i E_{J+1} t} |^2,
\end{equation}
from which Eq. (\ref{eq:main_answer}) straightforwardly follows.

\subsubsection{The physical (large N) limit: resummed function}
\label{sec:large}

To derive Eq. (\ref{eq:final_beauty}) we first elaborate
$\eta(N,J)$, Eq. (\ref{eq:coeff}), in the limit $N\gg 1$.  We
approximate the factorials using the Stirling formula: $z!\simeq
\sqrt{2\pi}e^{(z-1/2)\ln z -z}$.  As a result, an exponential
factor appears in the expression of $\eta(N,J)$.  We expand this
exponential in powers of $J/N$ and keep only the leading terms.
This is justified because the various factors in Eq.
(\ref{eq:coeff}) are significantly different from zero only
for $J\ll N$. From the procedure described, we obtain:
\begin{equation}\label{eq:approx_a}
    \eta(N,J)\simeq 2 N^{-1} (1+J) \exp\left[-(J+1)^2/N \right]~.
\end{equation}
This function has a peak at $J=\sqrt{N/2}-1$ and has a
characteristic width $2\sigma\sim\sqrt{N}$. This fact provides an
\textit{a posteriori} justification for taking $J\ll N$ for large
$N$.

As a second step, we notice that, for $N\gg 1$  the Fourier sum in
Eq.~(\ref{eq:main_answer}) is well approximated by the integral:
\begin{eqnarray}\label{eq:sumtoint}
   \sum_{J=0}^{N-1} \eta(N,J)\cos[(2J+2)g t]&\simeq &
   \int_0^\infty dJ\; \frac{2}{N} J
   e^{-J^2/N}\cos(2 J gt)\nonumber\\
   &=&1-\sqrt{\pi N} gt e^{-N g^2 t^2} {\rm erfi(\sqrt{N} g t)},
\end{eqnarray}
where the expression (\ref{eq:approx_a}) for $\eta$ has been used.
The function ${\rm erfi}(z)$ is given in Eq. (\ref{erfi}). From
Eq. (\ref{eq:sumtoint}) and (\ref{eq:main_answer}) the final result
(\ref{eq:final_beauty}) follows immediately.

\section{Discussion}
\label{sec:disc}

1. Let us compare the main features of our solution,
Eqs.~(\ref{eq:main_answer}) and (\ref{eq:final_beauty}), with the
corresponding results for a gas of small neutrino wavepackets. In
the latter case the survival probability exhibits a different
dependence on time, characterized by a simple exponential form
with the argument linear in $t$. To see this, consider a thought
experiment in which a beam of $2N$ electron neutrinos enters the
medium containing an equal mixture of electron and muon neutrinos.
It is straightforward to see that, for incoherent forward
scattering, the flavor composition of the beam will change
according to
\begin{eqnarray}
  dN_e/dt &=& -\lambda (N_e-N_\mu), \\
  dN_\mu/dt &=& -\lambda (N_\mu-N_e),
\end{eqnarray}
where $\lambda$ denotes the frequency of the flavor changing
collisions for a given neutrino. The solution is
\begin{eqnarray}\label{eq:equil_wp}
  N_e(t) &=& N(1+\exp(-\lambda t)),
\end{eqnarray}
a simple exponential. The origin of the difference between Eq.
(\ref{eq:final_beauty}) and (\ref{eq:equil_wp}) lies in the
different physics of the interaction. In the case of the
wavepackets, the time elapsed from $t=0$ tells us how many
background neutrinos a given neutrino interacted with; the
interaction time between neutrino pairs is very short. On the
other hand, in the case of the interacting plane waves, any two
waves in the box continuously interact with each other starting
from $t=0$. The scattering amplitude (and the effective
interaction cross section) grows with time. If the conversion
process were coherent, the survival probability would not depend
on the particular model adopted (see Sect.
\ref{subsect:reviewbell}).

2. It is interesting to note that the Fourier series in
Eq.~(\ref{eq:main_answer}) describes a periodic function with
period $T=\pi/g$.
This periodicity is lost when a transition to the integral is
made. Indeed, in Eq.  (\ref{eq:sumtoint}) the lowest frequency
over which the integration is performed is zero ($J=0$),
corresponding to infinite period.  It is clear that the period $T$
is much larger than the equilibration time $t_{eq}$ (see Eq.
(\ref{teq})): $T=\pi \sqrt{N} t_{eq}$.  Moreover, one can see that
typically $T$ exceeds the age of the universe: for a volume $V=1
~{\rm cm^{3}}$ we find $T \sim 10^{22}$ s. Therefore the integral
form (\ref{eq:final_beauty}) is an accurate approximation of the
exact result over time scales (at least) of the order of $t_{eq}$,
and applies to all the realistic physical situations.
\\

3. Let us comment on the range of validity of our results and their
possible extensions.

(a) Due to the $SU(2)$ invariance of the problem, the results are
the same if we replace the flavor eigenstates $\nu_e,\nu_\mu$ with
any pair of states in the flavor space, $\nu_x,\nu_y$, related to
the flavor basis by an $SU(2)$ transformation. Explicitly, if we
take neutrinos initially in the states $\nu_x$ and $\nu_y$, the
probability of conversion $\nu_x \rightarrow \nu_y$ does not
exhibit coherent effects in the large $N$ limit.  This tells us
that no coherent conversion is realized if coherence is initially
absent in the system, i.e. if, for a given neutrino $\nu_i$ being
in the flavor state $\nu_x$, the other neutrinos $\nu_j$ in the
ensemble are \emph{not} coherent superpositions of $\nu_x$ and
$\nu_y$: $\langle \nu_i | \nu_j \rangle = 0$ or $|\langle \nu_i |
\nu_j \rangle| = 1$.

(b) Several generalizations of our calculation are possible.  For
instance, one could investigate how the results depend on the
initial configuration of the system.  In \cite{us} we have studied
the case in which coherence is initially present in the system
($|\langle \nu_i | \nu_j \rangle| < 1$) and/or the initial state
is entangled. The findings were in agreement with the one-particle
description, Eq. (\ref{eq:oureqn}).

(c) The assumption of constant spin-spin coupling, $f_{ij}=1$,
could be relaxed to describe the case of, e.g., randomly
distributed couplings or interactions which depend on the
spin-spin distance or on the specific geometry of the system.
These cases may be difficult or impossible to describe
analytically; however we expect them to be characterized by the
same incoherent character of the conversion.

(d) The generalization to include vacuum oscillations and
interaction with ordinary matter can done by adding the
appropriate (well known) terms to the Hamiltonian.

\section{Conclusions}
\label{sec:conclusions}

We have found that a system of many ($N\gg 1$)
(forward-)interacting neutrino plane waves in a box reaches flavor
equilibration after a time $t\sim t_{eq}=(\sqrt{2} G_F
\sqrt{N}/V)^{-1}$ if the neutrinos are initially in flavor states.
This result agrees with what is expected from incoherent
scattering for this system, and therefore indicates the absence of
coherent conversion effects.

Although our results were obtained under specific simplifying
conditions (plane-waves in a box, constant couplings, forward
scattering only, etc.), they can be used to draw important general
conclusions on neutrino conversion in a neutrino background.  The
absence of coherent conversion has general character, since it does
not depend on the size of the neutrino wavepackets (see Sect.
\ref{subsect:singlevsmany}). It contrasts with ref.  \cite{Belletal}
and agrees with the prediction of the traditional -- and widely
applied -- single-particle description.  This tells us that there is
no reason to believe in a breakdown of this description and that,
therefore, its applications have to be considered valid. The
significance of this statement is clear in consideration of the
important physical problems, in cosmology and in stellar physics, in
which the neutrino-neutrino coherent scattering plays a role.

Finally, we remark that, in principle, the analytical solution we
have found for a system of interacting spins may be interesting on
its own and may find applications beyond the neutrino evolution
problem considered here.


\acknowledgments

We are grateful to G.~Raffelt for valuable comments.  A. F. was
supported by the Department of Energy, under contract
W-7405-ENG-36. A. F. is a Feynman Fellow at LANL. C. L. was
supported by the Keck foundation and by the National Science
Foundation grant PHY-0070928.

\bibliography{nunu_reply}
\bibliographystyle{JHEP}

\end{document}